\documentclass[prl,twocolumn,showpacs]{revtex4}
\usepackage{graphicx}
\begin{document}
%%%%%%%%%%%%%%%%%%%%%%%%%%%%%%%%%%%%%%%%%%%%%%%%%%%%%%%%%%%%%%%%%%%%%%%%%%%%%%%
\title{Eigenstates Ignoring Regular and Chaotic Phase-Space Structures} 
\author{Lars Hufnagel}
\author{Roland Ketzmerick} 
\author{Marc-Felix Otto}
\author{Holger Schanz}
\affiliation{
 Max-Planck-Institut f\"ur Str\"omungsforschung und 
 Institut f{\"u}r Nichtlineare Dynamik der Universit{\"a}t G{\"o}ttingen, 
 Bunsenstra{\ss}e 10, D-37073 G\"ottingen, Germany
}
%\email{holger@chaos.gwdg.de}
\date{\today}
\begin{abstract}
  We report the failure of the semiclassical eigenfunction hypothesis if
  regular classical transport coexists with chaotic dynamics.  All
  eigenstates, instead of being restricted to either a regular island or the
  chaotic sea, ignore these classical phase-space structures. 
  We argue that this is true even in the semiclassical limit for extended
  systems with transporting regular islands such as the standard map with
  accelerator modes.
\end{abstract}
\maketitle
%%%%%%%%%%%%%%%%%%%%%%%%%%%%%%%%%%%%%%%%%%%%%%%%%%%%%%%%%%%%%%%%%%%%

For a complete description of a quantum system the knowledge of its
spectrum and eigenstates is necessary.  It is a long standing goal to obtain
as much information about these fundamental quantum mechanical
objects as possible from the properties of the corresponding classical
system~\cite{Ein17,Gut90,Haake}.  The semiclassical eigenfunction
hypothesis~\cite{Per73,Ber77} states that in the semiclassical limit almost
all eigenstates are either regular or chaotic, i.e. their phase-space
representations are concentrated on regions with either regular or chaotic
classical dynamics, as illustrated in Figs.~1b and 1c. Exceptions are of
measure zero in the semiclassical limit and occur, e.g., at avoided level crossings
where regular and chaotic states can hybridize.  The semiclassical
eigenfunction hypothesis is supported by convincing numerical and experimental
data and forms the basis of the present understanding of spectral and
dynamical properties of \textit{finite} systems with a mixed phase
space~\cite{BR84,BTU93,PR93c,HT93,K+00,nobel,S+01}.

We study \textit{extended} systems with transporting regular islands, i.e.,
with a chain of islands which are traversed sequentially in time (arrows in
Fig.~1a).  Transporting islands are ubiquitous in physical systems.  In
particular, they occur in spatially periodic systems with time periodic
driving, like Hamiltonian ratchets~\cite{S+01}, atom-optic
experiments~\cite{ST+01,H+01}, or in the paradigmatic model of quantum chaos -- the
standard map~\cite{Izr90}, where they are called accelerator modes.

%%%%%%%%%%%%%%%%%%%%%%%%%%%%%%%%%%%%%%%%%%%%%%%%%%%%%%%%%%%%%%%%%%%%%%
\begin{figure}[!hb]
\centerline{\includegraphics[scale=0.45,angle=0]{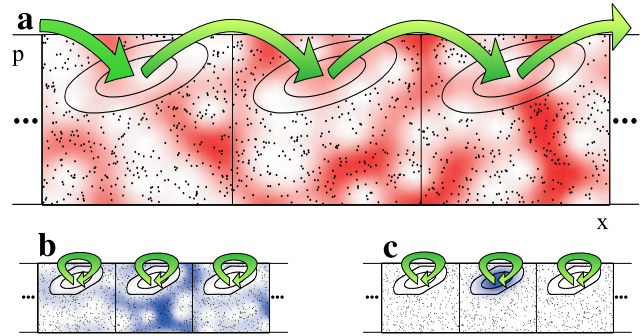}} 
\caption{ (color)
  Spatially periodic phase-space portraits (black dots) of the system given by
  Eq.(\ref{ham})~\cite{mapasym} showing a large island of regular motion and a
  chaotic sea in each spatial unit cell with $p \in [-1/2,1/2]$.  Eigenstates
  of the corresponding quantum system are projected onto phase space (colored
  density).  \textbf{(a)} The regular island is transporting as indicated by
  the arrows.  All eigenstates, like the one shown, spread over the regular
  islands as well as the chaotic sea. These "amphibious" eigenstates ignore
  the classical structures even though Planck's constant $h$ is smaller than
  the size of the island.  For comparison we show in (b) and (c) a system with
  regular islands of similar size that are not transporting~\cite{mapasym}.
  Two different types of eigenstates---\textbf{(b)} chaotic and \textbf{(c)}
  regular---exist, in agreement with the semiclassical eigenfunction
  hypothesis.  }
\end{figure}
%%%%%%%%%%%%%%%%%%%%%%%%%%%%%%%%%%%%%%%%%%%%%%%%%%%%%%%%%%%%%%%%%%%%%%
In view of the common confidence in the semiclassical eigenfunction hypothesis
and its importance, our result comes as a considerable surprise: The very
concept of regular and chaotic states does not apply in the presence of
transporting regular islands. Instead, we find that all eigenstates spread
over regular islands as well as the chaotic sea of classical phase space. An
example of such an "amphibious" eigenstate can be seen in Fig.~1a.  We argue
that even in the semiclassical limit \textit{all} eigenstates will ignore
classical phase-space structures as soon as transporting islands are present
and we will discuss a number of questions raised by this failure of the
semiclassical eigenfunction hypothesis for extended systems.

In order to allow for an efficient calculation of quantum eigenstates for
systems with transporting islands we consider one-dimensional kicked
Hamiltonians
\begin{equation}\label{ham}
H = T(p) + V(x) \sum_n \delta(t-n)
\end{equation}
with a classical time-evolution described by the map
$$
x_{t+1} = x_t + T'(p_{t}),\qquad
p_{t+1} = p_t - V'(x_{t+1})\,.
$$
We are interested in spatially periodic dynamics of period 1 and consider
the simple situation where also $T(p+1)=T(p)$ with a cyclic momentum variable,
$p\equiv p+1$. This ensures that the phase-space averaged current is zero. We
choose the functions $V'(x)$ and $T'(p)$ such that they give rise to a large
regular island transporting in $x$-direction~\cite{mapasym} as can be seen in
Fig.~1a. This makes the semiclassical regime, where the island area is
bigger than Planck's constant $h$, numerically more easily accessible.

The quantum mechanical time-evolution operator for one period,
\begin{equation}
\hat U|\phi(t)\rangle = |\phi(t+1)\rangle ,
\end{equation}
is given by $\hat U=\exp(-2\pi i T(\hat p)/h_{\text{eff}})\,\exp(-2\pi i
V(\hat x)/h_{\text{eff}})$, where $h_{\text{eff}}$ is dimensionless and
irrational, denoting the ratio of Planck's constant $h$ to the phase-space
area of a unit cell.  The irrationality of $h_{\text{eff}}$ makes this quantum
system aperiodic and excludes the applicability of Bloch's theorem which
otherwise would lead to spatially periodic eigenstates. We thereby model the
experimentally relevant situation of spatially periodic systems with some
disorder \cite{rational}.  The eigenstates $|\psi\rangle$ of $\hat U$,
\begin{equation}\label{ef}
\hat U|\psi\rangle=e^{-i \epsilon}|\psi\rangle ,
\end{equation}
are the objects of our interest.

These eigenstates can be represented in phase space (Husimi
representation~\cite{H+84a}) and are compared with the classical phase-space
portrait in Fig.~1.  In contrast to the semiclassical eigenfunction hypothesis
we find that all eigenstates are amphibious in all spatial unit cells, living
on regular islands as well as in the chaotic sea.  For comparison, we present
in Figs.~1b and 1c a system which is equivalent in terms of phase-space
portrait, island size, and effective Planck's constant $h_{\text{eff}}$, but
with non-transporting islands.  In this case the eigenstates can be
categorized as chaotic (Fig.~1b) or regular (Fig.~1c) in agreement with the
semiclassical eigenfunction hypothesis.  This demonstrates that the mere
presence of \textit{transporting} regular islands leads to a failure of the
semiclassical eigenfunction hypothesis.

%%%%%%%%%%%%%%%%%%%%%%%%%%%%%%%%%%%%%%%%%%%%%%%%%%%%%%%%%%%%%%%%%%%%%%
\begin{figure}[bt] 
\centerline{\includegraphics[scale=0.42,angle=0]{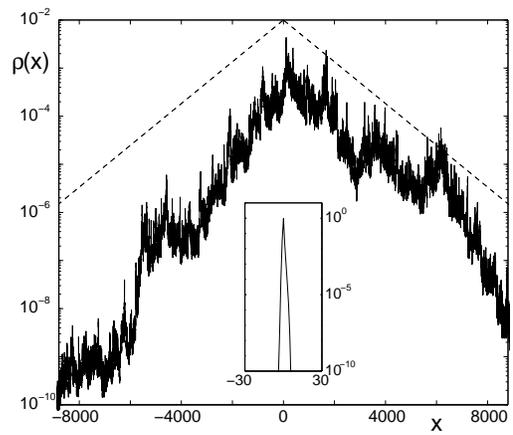}}
\caption{
  The weight of the amphibious eigenstate from Fig.~1a in each unit cell,
  $\rho(x) = \int_{x-1/2}^{x+1/2} dx' \; |\psi(x')|^2$, shows an overall
  exponential decay with strong fluctuations.  The localization length is
  several orders of magnitude bigger than in the case of the system of Fig.~1b
  with non-transporting islands (inset).  It is quite well estimated by
  $\exp(-|x|/\lambda)$ (dashed line), with $\lambda=v/\gamma$ determined from
  the tunneling rate $\gamma$ out of the regular island with velocity $v=1$.
  }
\end{figure}
%%%%%%%%%%%%%%%%%%%%%%%%%%%%%%%%%%%%%%%%%%%%%%%%%%%%%%%%%%%%%%%%%%%%%%

While Fig.~1 shows the local structure of the eigenstates on the scale of a
few unit cells, amphibious eigenstates are also distinguished from chaotic
eigenfunctions from a global perspective.  We find that they are localized
with a much larger localization length than without transporting islands
(Fig.~2), as anticipated from previous studies of wave packet
dynamics~\cite{H+84,SZ99,IZ00,IFZ02}.  This localization length of the
eigenfunctions is determined by the tunneling out of the island: A wave packet
initialized at the center of a regular island initially follows the classical
dynamics (Fig.~3). Its weight inside the island, however, decays as
$P(t)=\exp(-\gamma t)$, where $\gamma$ is the tunneling rate between the
regular island and the chaotic sea~\cite{H+84}.  During the time scale
$1/\gamma$ of this decay the wave packet is transported with the velocity
$v=1$ of the island.  We thus estimate the localization length as $\lambda =
v/\gamma$.  We determined the tunneling rate $\gamma=9.885\:10^{-4}$ from the
wave packet dynamics.  The localization length of the amphibious eigenstates
agrees well with the length $\lambda=1011$ estimated from this tunneling rate,
as can be seen in Fig.~2.  The scaling of the tunneling rate $\gamma \sim
\exp(-C/h_{\text{eff}})$~\cite{H+84}, with some constant $C$, yields a
localization length $\lambda \sim \exp(C/h_{\text{eff}})$ that increases
exponentially with decreasing $h_{\text{eff}}$. As a numerical computation of
eigenstates requires a system size of several localization lengths, this
scaling implies that approaching the semiclassical limit of small
$h_{\text{eff}}$ in the presence of transporting islands entails an enormous
increase in numerical effort.  Note that in the non-transporting case, $v=0$,
the above estimate for the localization length breaks down and the usual
dynamical localization with $\lambda \sim h_{\text{eff}}^{-1}$ takes place.

%%%%%%%%%%%%%%%%%%%%%%%%%%%%%%%%%%%%%%%%%%%%%%%%%%%%%%%%%%%%%%%%%%%%%%
\begin{figure}[bt] 
\centerline{\includegraphics[scale=0.42,angle=0]{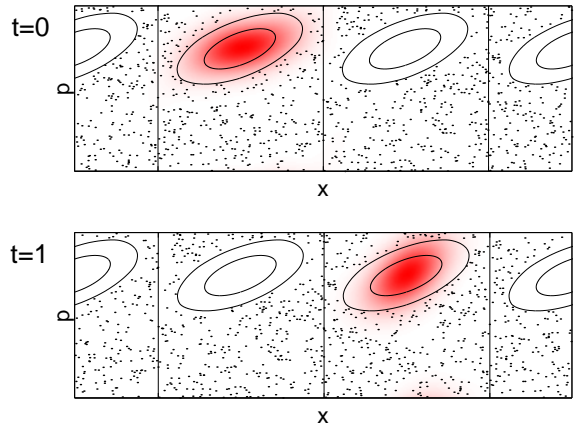}}
\caption{ Quantum evolution of a wave packet (gray shade) from $t=0$ (top) to
$t=1$ (bottom) for the system of Fig.~1a where classically a regular island
moves one unit cell to the right in one time period.  Although the underlying
amphibious eigenstates ignore the classical phase-space structure they
conspire such that the initial dynamics of wave packets follows the
semiclassical expectation. }
\end{figure}
%%%%%%%%%%%%%%%%%%%%%%%%%%%%%%%%%%%%%%%%%%%%%%%%%%%%%%%%%%%%%%%%%%%%%%
An intuitive understanding of why transporting islands lead to amphibious
eigenstates can be gained in the following way: The quantum mechanical
continuity equation of a state $\varphi$ is
\begin{equation}
\frac{\partial}{\partial t} |\varphi(x,t)|^2 
+ \frac{\partial}{\partial x} J_{\varphi}(x,t) = 0\,,
\end{equation}
where $J_\varphi$ is the probability current of $\varphi$~\cite{discrete}.
Integration over one temporal period leads to
\begin{equation}
|\varphi(x,t+1)|^2 - |\varphi(x,t)|^2 = - \frac{\partial}{\partial x}
\int_t^{t+1}dt J_{\varphi}(x,t)\,.
\end{equation}
The left-hand side vanishes for eigenstates $|\psi\rangle$ of the
time-evolution operator over one period, Eq.~(\ref{ef}).  Thus their temporally
averaged current is spatially constant.  Moreover, this constant is zero,
because all eigenstates are localized and thus have a vanishing overall
current.  Therefore, the probability current
\begin{equation}
\label{current}
\int_t^{t+1}dt J_{\psi}(x,t) \equiv 0 ,
\end{equation}
vanishes for all $x$. We can use this condition to infer the local structure
of the eigenstates in the semiclassical regime, where the quantum probability
current $J_{\psi}$ is close to the classical current.  In particular, we can
immediately exclude regular eigenstates that are mainly concentrated on the
transporting islands and thus would show a considerable current. Similarly,
chaotic eigenstates can be excluded, as the chaotic sea also shows a non-zero
current in the opposite direction~\cite{S+01}. The amphibious eigenstate
presented in Fig.~1a, however, is distributed uniformly over the phase-space
regions with opposing classical currents. It fulfills the requirement of a
vanishing current of Eq.~(\ref{current}) because the phase space averaged
classical current is zero.  Hence we can conclude that even in the
semiclassical limit \textit{all} eigenstates spread simultaneously over
regular and chaotic regions of the phase space in all spatial unit cells.

It is illuminating to point out where the arguments leading to the
semiclassical eigenfunction hypothesis fail in the case of transporting
islands.  In the semiclassical limit the tunneling rate $\gamma$ between the
center of a regular island and the chaotic sea decreases to arbitrarily small
values. Perturbation theory in $\gamma$ can be used to calculate the
eigenstates provided that the dimensionless coupling is small,
$\gamma/\Delta\ll 1$, where $\Delta$ denotes the mean level spacing of those
unperturbed states that will be coupled by tunneling. In extended systems it
is crucial to understand the dependence of $\Delta$ on the system size $L$: In
the absence of classical transport the unperturbed regular and chaotic states
are localized, and the tunneling out of the island may couple only groups of
states that are spatially close.  In particular, the number of these states
and their mean level spacing is fixed for given $h_{\text{eff}}$ and does not
depend on the system size, leading to $\Delta\sim h_{\text{eff}}$.  Using
$\gamma \sim \exp(-C/h_{\text{eff}})$ we see that perturbation theory is
applicable $(\gamma/\Delta \ll 1)$ for small $h_{\text{eff}}$.  Hence the
semiclassical eigenfunction hypothesis is valid.  In the presence of
transporting islands, however, the unperturbed regular states are extended and
all of them will be coupled to all chaotic states.  The mean level spacing of
the unperturbed states is now inversely proportional to the system size,
$\Delta\sim h_{\text{eff}} L^{-1}$, and perturbation theory breaks down
$(\gamma/\Delta \approx 1)$ as soon as $L\sim h_{\text{eff}}/\gamma$. In
particular, we conclude that for an infinite system, transporting regular and
chaotic regions are strongly coupled even in the semiclassical limit
\cite{Thou}. From the breakdown of perturbation theory alone, however, we
cannot conclude the locally amphibious nature of eigenstates which we find
numerically and deduce from the condition of vanishing current Eq.~(\ref{current}).

%%%%%%%%%%%%%%%%%%%%%%%%%%%%%%%%%%%%%%%%%%%%%%%%%%%%%%%%%%%%%%%%%%%%
\begin{figure}[hbtp] 
\centerline{\includegraphics[scale=0.48,angle=0]{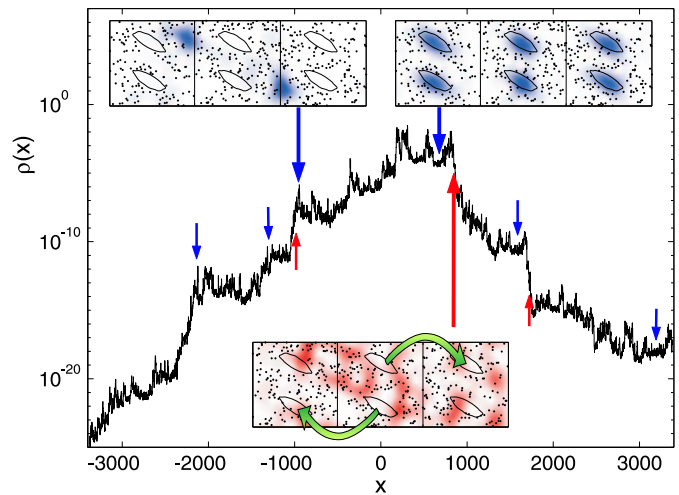}}
\caption{ (color) A typical eigenstate for a system with a symmetric phase
space and islands transporting in opposite directions (arrows in lower
inset)~\cite{mapsym}.  Several regions are projected onto phase space (insets)
demonstrating the mixed type of this eigenstate, with some regions being
mainly regular, some mainly chaotic, and some amphibious.  Each type appears
at many locations of the eigenstate (short arrows).  }
\end{figure}
%%%%%%%%%%%%%%%%%%%%%%%%%%%%%%%%%%%%%%%%%%%%%%%%%%%%%%%%%%%%%%%%%%%%

What do eigenstates look like if one has several transporting islands with
different velocities?  Let us illustrate this question with the important case
of a symmetric phase space containing equivalent islands transporting in
opposite directions (arrows in the lower inset of Fig.~4), as in the standard
map with accelerator modes~\cite{Izr90}.  In this case regular eigenstates that are supported
simultaneously by both islands would fulfill the condition of vanishing
current, Eq.~(\ref{current}).  However, due to the breakdown of the
perturbation theory, we do not expect such states.  A numerical investigation
of the eigenstates reveals a structure which is consistent with all our
arguments but surprisingly complicated: We find that all eigenstates are of a
mixed type -- in some unit cells they are concentrated on both regular
islands, in some they spread over the chaotic sea, and in some unit cells they
are amphibious (Fig.~4). This unexpected finding calls for further
exploration.

Although the amphibious eigenstates disregard the classical phase-space
structure, our results do not contradict quantum-classical correspondence.
For example, the wave packet of Fig.~3 follows the classical expectation for
arbitrarily long times $t\to\infty$ , if the limit $h_{\text{eff}}\to 0$ is
taken first.  It is well known that eigenstates, where the order of these
limits is reversed, might differ from the semiclassically expected states,
which are called quasimodes \cite{Arn72,Ber77a}.  A famous example is the
symmetric double-well potential, where eigenstates must be symmetric or
antisymmetric and thus live in both wells, while classically the dynamics is
restricted to one of the wells.  This is a very special situation, however, as
quasimodes and eigenstates will again coincide if the symmetry of the
double-well potential is weakly perturbed.  Amphibious eigenstates are a
non-trivial example for the difference between quasimodes and eigenstates.
Moreover, for extended systems with transporting islands they are a generic
phenomenon.

The failure of the semiclassical eigenfunction hypothesis and the appearance
of amphibious eigenstates raise further questions: transporting island chains
of arbitrary length exist even in systems with a finite phase space, such as
in the neighborhood of the boundary circle enclosing a regular island.  Will
there be amphibious eigenstates?  The statistical properties of chaotic
eigenstates~\cite{Haake} can be seen in conductance measurements on quantum
dots~\cite{Alh00}.  What are the statistical properties of amphibious
eigenstates?  Are they similar to those of chaotic eigenstates or will there
be specific correlations caused by the regular islands? 

The recently developed techniques to observe atom dynamics in optical 
lattices~\cite{ST+01,H+01} are perfectly suited for the experimental study of
amphibious eigenstates.  In such systems wave packets can be prepared on
selected points in phase space, e.g. on the center of an island, and their
long-time dynamics can be studied.  A measurement of the phase-space
distribution would reveal that the asymptotic wave packet is uniformly
distributed over phase space, independent of the initial preparation. This is
in sharp contrast to a system with non-transporting islands and would be a
clear experimental signature of amphibious eigenstates.

We acknowledge discussions with T.~Dittrich, D.~Cohen, S.~Fishman, A.~Iomin, and F.~Wolf.
%%%%%%%%%%%%%%%%%%%%%%%%%%%%%%%%%%%%%%%%%%%%%%%%%%%%%%%%%%%%%%%%%%%%
%\bibliography{paper}

\begin{thebibliography}{10}
\bibitem{mapasym}
\bibinfo{note}{In order to get large regular islands we start from piecewise
  linear functions $t'(p)=s/4\pm(A-sp)$ for $0<\pm p<1/2$ (periodically
  extended beyond the unit cell) and $v'(x) = -m-r(x-m)$ for $m-1/2<x<m+1/2$,
  where $m$ is an integer. Quantum anomalies are avoided by smoothing these
  discontinuous functions $t'$ and $v'$ with a Gaussian,
  $G(z)=\exp(-z^2/2\varepsilon^2)/ \sqrt{2\pi\varepsilon^2}$, finally yielding
  analytic functions $T'(p) = \int dz \; t'(p+z) \; G(z)$ and $V'(x) = \int dz
  \; v'(x+z) \; G(z)$. For a transporting island we choose parameters $A=1$,
  $s=2$, $r=0.65$, and $\varepsilon=0.015$, while for a non-transporting island
  of roughly the same size we take $A=0$, $s=2$, $r=0.65$, and
  $\varepsilon=0.11$. The islands obtained in this way are approximately 50
  times larger than the largest accelerator modes of the standard map. For
  quantum calculations in these systems we use $h_{\rm eff}=1/(9+\sigma)$,
  where $\sigma=(\sqrt{5}-1)/2 \approx 0.618$, such that the island size is
  roughly $2h$.}
\bibitem{discrete}
\bibinfo{note}{For notational simplicity we treat $x$ as a continuous variable
  even though, due to the systems periodicity in $p$, it is quantum
  mechanically discrete.}
\bibitem{mapsym} \bibinfo{note}{The system is given by Eq.(\ref{ham}) with
$T'(p) = 1.22 \sin 2\pi p + 0.2 \sin 4\pi p + 0.15 \sin 6\pi p$ and
$V'(x)=-x$. It is a modification of the standard map ($T'(p)=p$ and
$V'(x)=-K/(2\pi)\sin 2\pi x$) that enlarges the transporting island by a
factor of 25 compared to the largest accelerator mode. In addition, we
exchange $p$ and $x$.  For quantum calculations we use $h_{\rm
eff}=1/(20+\sigma)$ such that the island size is roughly $2h$.}
\end{thebibliography}

%%%%%%%%%%%%%%%%%%%%%%%%%%%%%%%%%%%%%%%%%%%%%%%%%%%%%%%%%%%%%%%%%%%%
\end{document}